\documentclass[a4paper]{spie}  
\usepackage[english]{babel}
\usepackage[top=2.54cm, bottom=4.94cm, left=1.925cm, right=1.925cm]{geometry}
\usepackage{amsmath}
\usepackage{graphicx}
\usepackage{caption}
\usepackage{subcaption}
\usepackage{pbox}
\usepackage[colorlinks=true, allcolors=blue]{hyperref}
\usepackage{titlesec}
\usepackage{enumitem}
\usepackage{caption}
\usepackage{csquotes}
\usepackage{algorithm}
\usepackage{algpseudocode}
\usepackage{multirow}
\usepackage{graphicx, caption}
\usepackage{array}
\newcolumntype{?}{!{\vrule width 1pt}}
\setlist{nolistsep}
\usepackage{graphicx}
\newsavebox{\imagebox}
\usepackage{color,soul}
\usepackage{tabularx}
 \usepackage{ctable}
\usepackage{amsmath,amsfonts,amssymb}
\usepackage{graphicx}
\usepackage[colorlinks=true, allcolors=blue]{hyperref}
\usepackage{ctable}

\title{Data variation-aware medical image segmentation}

\author[a]{Arkadiy Dushatskiy}
\author[b]{Gerry Lowe}
\author[a,~c]{Peter A. N. Bosman}
\author[d]{Tanja Alderliesten}

\affil[a]{Centrum Wiskunde \& Informatica, Science Park 123,
1098XG, Amsterdam, the Netherlands}
\affil[b]{Mount Vernon Cancer Centre, Rickmansworth Rd, HA62RN, Northwood, the United Kingdom}
\affil[c]{Delft University of Technology, Van Mourik Broekmanweg 6,
2628XE, Delft, the Netherlands}
\affil[d]{Leiden University Medical Center, Albinusdreef 2,
2333ZA, Leiden, the Netherlands}

\authorinfo{Further author information: (Send correspondence to A. Dushatskiy)\\A. Dushatskiy: E-mail: arkadiy.dushatskiy@cwi.nl}

 \makeatletter
\def\thickhline{%
  \noalign{\ifnum0=`}\fi\hrule \@height \thickarrayrulewidth \futurelet
   \reserved@a\@xthickhline}
\def\@xthickhline{\ifx\reserved@a\thickhline
               \vskip\doublerulesep
               \vskip-\thickarrayrulewidth
             \fi
      \ifnum0=`{\fi}}
\makeatother

\newlength{\thickarrayrulewidth}
\begin{document} 
\maketitle

\renewcommand{\labelenumii}{\theenumii}
\renewcommand{\theenumii}{\theenumi.\arabic{enumii}.}

\begin{abstract}
Deep learning algorithms have become the golden standard for segmentation of medical imaging data. In most works, the variability and heterogeneity of real clinical data is acknowledged to still be a problem. One way to automatically overcome this is to capture and exploit this variation explicitly. Here, we propose an approach that improves on our previous work in this area and explain how it potentially can improve clinical acceptance of (semi-)automatic segmentation methods.

In contrast to a standard neural network that produces one segmentation, we propose to use a multi-path Unet network that produces multiple segmentation variants, presumably
corresponding to the variations that reside in the dataset. Different paths of the network are trained on disjoint data subsets. Because a priori it may be unclear what variations exist in the data, the subsets should be automatically determined. This is achieved by searching for the best data partitioning with an evolutionary optimization algorithm. Because each network path can become more specialized when trained on a more homogeneous data subset, better segmentation quality can be achieved. In practical usage, various automatically produced segmentations can be presented to a medical expert, from which the preferred segmentation can be selected. In experiments with a real clinical dataset of CT scans with prostate segmentations, our approach provides an improvement of several percentage points in terms of Dice and surface Dice coefficients compared to when all network paths are trained on all training data. Noticeably, the largest improvement occurs in the upper part of the prostate that is known to be most prone to inter-observer segmentation variation.
\end{abstract}


\section{INTRODUCTION}
This work is focused on (semi-)automatic segmentation of medical images with deep learning algorithms while capturing data variation. Fundamentally, our approach can detect and tackle any existing source of variation in the data, e.g., different scanning devices or different patient groups. Specifically, we focus on observer segmentation variation and, particularly, consider the following scenario: (1) In a dataset there are multiple scans, one scan per patient. (2) Each scan is segmented by one observer, the name of the observer is normally unknown. In contrast to other works which focus on medical image segmentation taking into account the observer variation \cite{kohl2018probabilistic, rupprecht2017learning}, our goal is not to model the data uncertainty and use a probabilistic model to produce a set of possible segmentations, but to explicitly capture observer variation, and produce segmentation variants corresponding to different ways of segmenting residing in the dataset. Moreover, we do not assume that for each scan segmentations by mutiple observers are available. We believe such a situation is quite common in clinical practice.

We envision a clinical usage scenario as follows. For each scan that needs to be segmented (e.g., for a brachytherapy treatment planning), our neural network (previously trained) produces multiple organ segmentation variants. A medical expert can choose one of those for further clinical usage. Additional manual adjustments are possible but much less likely needed since there were multiple options in-line with the original variation in the data to choose from. Thus, the goal of the neural network is to produce segmentations such that at least one of them is close to the expert's segmentation understanding for that particular scan. We propose an approach to get more accurate and more diverse segmentation variants by using different data subsets during training. The best dataset partitioning is found automatically using an efficient optimization algorithm.
Such an approach has two major goals: (1) Improve the segmentation quality as each net is trained on more homogeneous data and thereby more accurately learn each possible specific segmentation variant rather than learning just one, most probably an average, segmentation. (2) Multiple variants of segmentation can be presented and it is more likely that an observer agrees with one of the variants than with an average segmentation. These two points may well contribute to the ultimate goal of our work - increased acceptance and uptake of (semi-)automatic segmentations in clinical practice.

To the best of our knowledge, the only similar approach was initially proposed by us in \cite{dushatskiy2020observer}. However, in our previous work substantially large simulated segmentation variations were used, and the approach was mainly suited for only two types of variations. Here, we propose an efficient algorithm capable of capturing an arbitrary number of variations (defined by the user beforehand) and apply it to segmentation variations present in real clinical data.

\section{METHOD}
\subsection{Multi-path segmentation networks}
Ultimately, in our approach, each segmentation variant is supposed to be learned by a separate segmentation network. In common Unet-like segmentation architectures  \cite{unet}, which consist of an encoder and a decoder, an encoder learns compressed representations of image features, while the decoder translates them to a segmentation prediction. It is reasonable to utilize the same encoder for different segmentation variants, while connecting it to muiltiple decoders because (1) We can expect that the image features required to properly predict variants are common. (2) A shared encoder improves computational efficiency.

We assume that our segmentation network consists of an encoder $E$ and multiple decoders $D_1,\dots,D_{\alpha}$. The $i$-th segmentation variant for input image $x$ is produced as $D_i(E(x))$. The architecture scheme is shown in Figure~\ref{figure:unet}. 
In each training epoch, each decoder is trained using the corresponding training data subset $P_i$. Note that the encoder is trained using the full dataset $P$. This training procedure is shown in pseudocode in Algorithm~\ref{algo:algorithm1}.

\subsection{Optimization procedure} \label{sub:optim}
We formulate the problem of partitioning a set of scans $P$ ($|P|$=N) into $\alpha$ disjoint subsets $P_1, P_2, \dots ,P_\alpha$: $P_1 \cup P_2 \dots \cup P_\alpha= P, P_1 \cap P_2 \dots \cap P_\alpha= \emptyset$ as an optimization problem. The search space (space of solutions) is the space of vectors of length $N$ over alphabet $\{1,2,\dots,\alpha\}$, i.e., $\{1,2,\dots,\alpha\}^N$. These vectors determine the partitioning: a value $k$ in position i determines that the i-th scan belongs to $P_k$.
For optimization we use a recently introduced efficient surrogate-assisted evolutionary algorithm \emph{SA-P3-GOMEA} \cite{dushatskiy2021novel} that was specifically designed for discrete optimization problems with expensive function evaluations. In this optimization problem, one function evaluation means a full training and evaluation of a network with a suggested dataset partitioning. As this is a computationally expensive procedure, an efficient optimization algorithm might be used to reduce the overall computational costs. 

To define the function to be optimized, resulting in the best partitioning, we explain how each partitioning is evaluated. After a net is trained using Algorithm~\ref{algo:algorithm1}, we use a fixed validation set $V$ for evaluation. For each scan from $V$, the trained network produces $\alpha$ segmentation variants. For each variant, its quality is computed using a segmentation performance metric. The variant with the best score, is selected for that scan. The final validation score is averaged over all scans in $V$. This score is the optimized (maximized) function value. Such an evaluation procedure simulates a usage scenario, such, that for each scan a medical expert is shown multiple segmentation variants and is interested in selecting the best one. To avoid overfitting on the validation set, we use an additional test set, which is left out during the optimization and is used only for obtaining the final results.

\subsection{Segmentation quality evaluation} \label{sub:evaluation}
We consider common segmentation performance metrics: the Dice-S\o rensen Coefficient (DSC), and the Surface Dice-S\o rensen Coefficient \cite{surfacedice} (SDSC). While the DSC measure considers only global segmentation quality accuracy without focusing on the accuracy near organ borders, the SDSC indicates the ratio of the contour that deviates from the reference by no more than $\tau$ mm (the tolerance value $\tau$ is a chosen constant). An SDSC value of one means no manual adjustment is needed if the acceptable deviation from the reference is $\tau$ mm. Inside the optimization procedure we use the average of DSC and SDSC (with $\tau=2mm$) to combine both global and local segmentation quality metric values.

\subsection{Segmentation neural network architecture} \label{sub:networks}
Our approach is not limited to a particular segmentation network architecture. However, due to the computational complexity (many trainings inside the optimization loop), the used segmentation network has to be computationally efficient. Thus, we selected two simple, yet efficient (found in preliminary experiments) Unet architectures from \cite{Yakubovskiy:2019}. The selected networks differ in the implemented encoder network. The first one is based on the ResNet18 \cite{he2016deep}, while the second one has the VGG19 \cite{SimonyanZ14a} encoder. Preliminary experiments demonstrated that VGG19-Unet shows better performance than ResNet18-Unet at the cost of a longer training time. While ultimately we are interested in using the best performing network, we show results on both selected networks in order to demonstrate the independence of our approach from a particular architecture.

\begin{minipage}{0.45\textwidth}
\begin{algorithm}[H]
    \centering
    \caption{Training routine for the proposed segmentation approach}\label{algo:algorithm1}
    \footnotesize
   \hspace*{\algorithmicindent}  \textbf{Input:} the number of demanded segmentation variants $\alpha$; dataset partitioning {$P_1,\dots,P_\alpha$}; architecture encoder $E$, decoders $D_1,\dots,D_\alpha$, the number of training epochs $nEpochs$ \\
    \Comment{$X$ denotes an image, $y$ is a corresponding reference segmentation, $\hat{y}$ is a predicted segmentation}
        
    \begin{algorithmic}[1]
         \For{$epoch\gets 1, nEpochs$}
        \For{$i \in randomPermutation([1,\dots,\alpha])$ }
        \For{$\text{batch of pairs} (X, y)  \in P_i$} 
            \State $\hat{y} \gets E(D_i(batch))$
            \State $L \gets loss(y,\hat{y})$
            \State update weights of $E,D_i$ by backpropagation using $L$
        \EndFor
        \EndFor
        \EndFor
    \end{algorithmic}
\end{algorithm}
\hspace{1cm}
\end{minipage}
\hspace{1.5cm}
\begin{minipage}{0.4\textwidth}
\centering
    \includegraphics[width=1.0\textwidth]{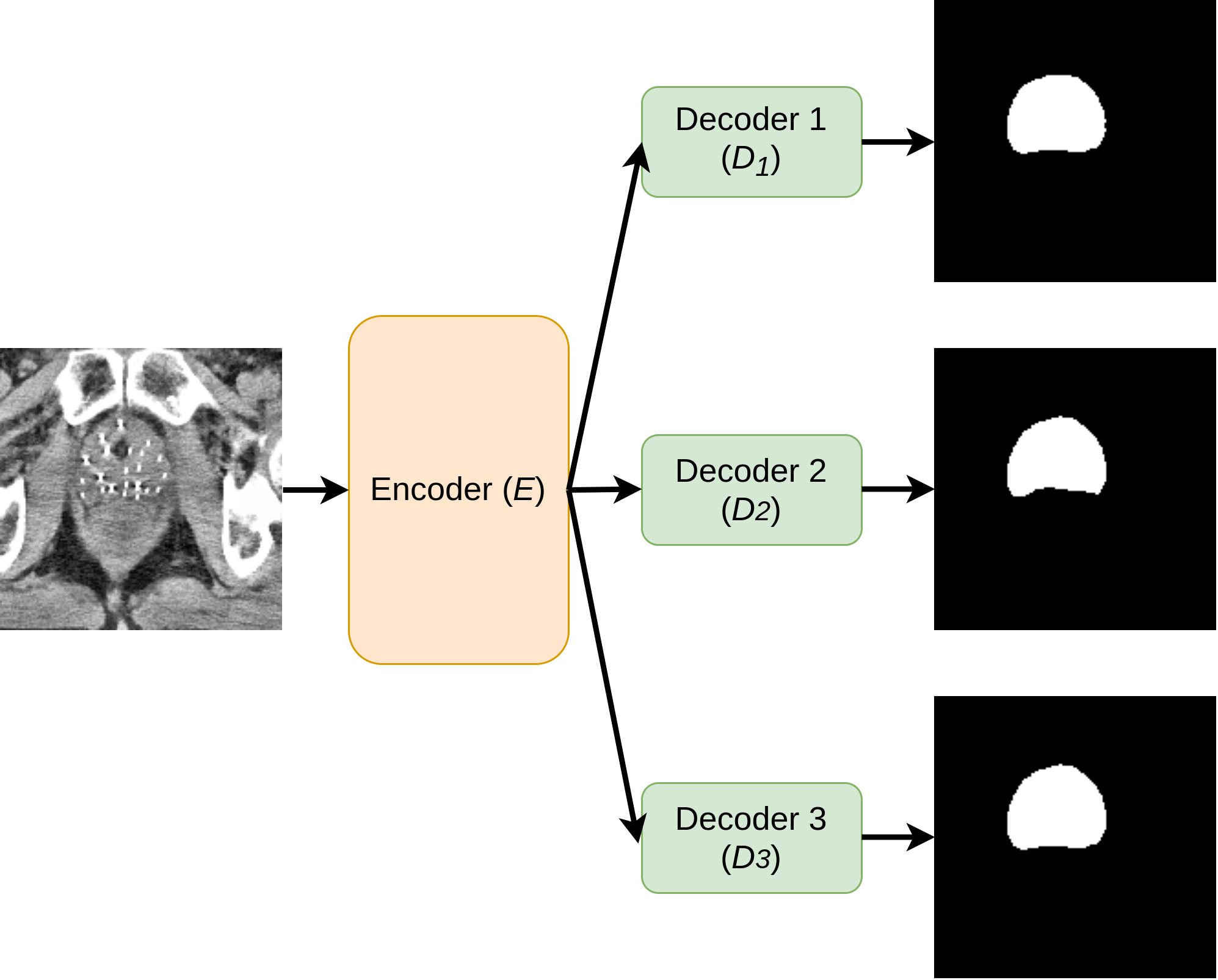}
    \captionof{figure}{A schematic illustration of the used multi-path Unet network. Multiple decoders (3 in this case) are attached to the shared encoder and produce three different segmentation predictions.}
    \label{figure:unet}
\end{minipage}

\section{EXPERIMENTS}
\subsection{Data}
We use a real clinical dataset from Mount Vernon Cancer Centre (Northwood, UK) of 196 prostate cancer patients treated with external beam radiation treatment and brachytherapy. For each patient, we have one CT scan with the corresponding organ segmentations that were used for brachytherapy treatment planning purposes. In the preliminary preprocessing step, 13 scans with severe visual artefacts are filtered out. From the remaining 183 scans, 40 and 43 scans were left out for validation and test sets respectively. The remaining 100 scans can be used for partitioning optimization and training. We consider a segmentation task with two classes: prostate and background. The initial $3D$ data is transformed to $2D$ slices. Each scan consists of 18 slices on average (range: 10 to 27 slices). In order to balance the number of background and prostate pixels, each slice is cropped to the $150\text{mm} \times 150\text{mm}$ central region of the image. The pixel spacing of the processed slices is $1.17\text{mm} \times 1.17\text{mm}$. 


\subsection{Experimental setup} \label{sub:setups}
To show that our approach works with datasets of different sizes, we perform experiments with 50 and 100 patients in the training part ($N=50, 100$). We set the number of evaluated partitionings in each optimization run in case of $N=50$ to $2000$. This value was chosen as a tradeoff between computational time and quality of the found partitionings. As one epoch of training with 100 scans is approximately two times longer than the epoch of training with 50 scans, we set the number of evaluated partitionings in case of $N=100$ to $1000$. We study the results when various numbers of partitions are used: $\alpha=2,3,4$. As an additional baseline, we add a standard single-path network ($\alpha=1$). As both the neural network training and the used optimization algorithm are naturally stochastic, we repeat an experiment three times for each value of $\alpha$, and each used network architecture (ResNet18-Unet or VGG19-Unet). All results are reported on the test set which guarantees no overfitting (on the validation set). The reported values are averaged over the three runs.

Our experiments aim at showing that our approach results in better segmentation performance in comparison to the standard training approach, i.e., with no data partitioning (all network paths are trained on the full training dataset). 

\section{RESULTS} \label{sub:results}
Main results are shown in Table~\ref{tab:results}. The proposed approach shows better performance than a corresponding multi-path network trained without dataset partitioning by approximately 1\% in DSC coefficient when $N=50$. More substantial differences of up to $5\%$ and $1.5\%$ can be observed in SDSC metric with $\tau=2mm$ and $\tau=4mm$ respectively. Note that the smaller $\tau$ value means less tolerance to the deviations from the reference segmentation. Furthermore, our approach with $\alpha$ values of 2, 3, and 4 shows better performance than a standard single-path network $\alpha = 1$. In case of $N=100$, the improvements of our approach are smaller ($1.5-2.5\%$ in SDSC, $\tau=2mm$, $0.5-1\%$ in DSC), though it still managed to find an improvement over the baselines in all cases. We hypothesize, that running optimization for longer might result in finding better partitionings, but such experiment was not conducted due to limited available computation resources. We separately study the effect of our approach on the segmentation quality in three parts of the prostate: the middle part (\emph{mid-gland}), the upper part (\emph{base}), and the lower part (\emph{apex}).
Finally, we visualize the obtained segmentations. 

\begin{figure}[h]
  \includegraphics[width=.9\textwidth]{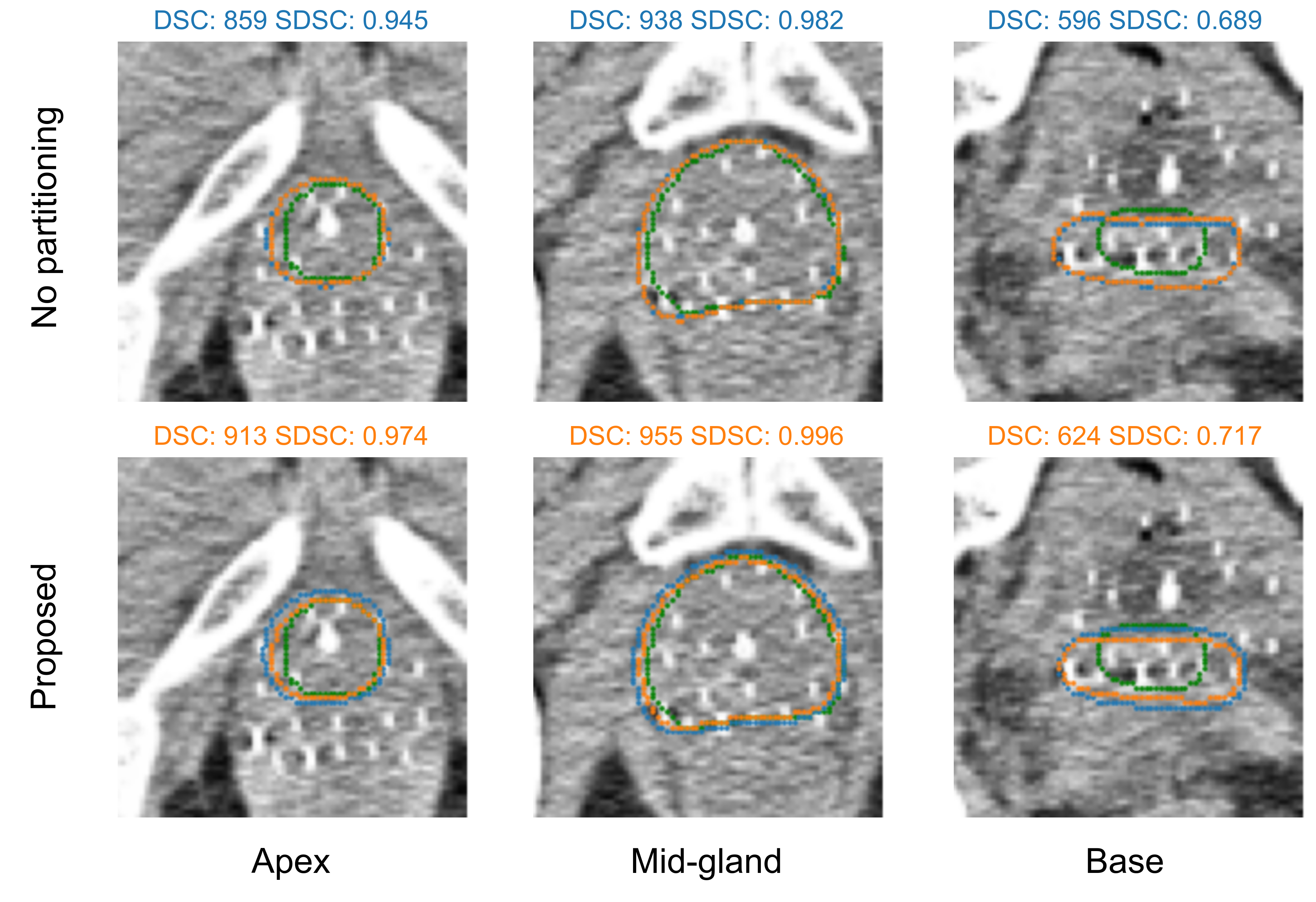}
  
    \caption{Results demonstration in case of $\alpha=2$ for one of the patients. The reference segmentation is shown in green, the orange and the blue ones show two produced segmentation variants. The upper row corresponds to the training with no partitioning (see Section~\ref{sub:results}), the bottom row corresponds to the training using the best found partitioning. Scores for the selected (best among the two variants) segmentations are showed with a corresponding color. The proposed method produces more accurate segmentations. All images are zoomed-in by a factor of two.}
\label{fig:pictures}
\end{figure}

\begin{figure}[h]
\vspace{0.5cm}
    \includegraphics[width=0.9\textwidth]{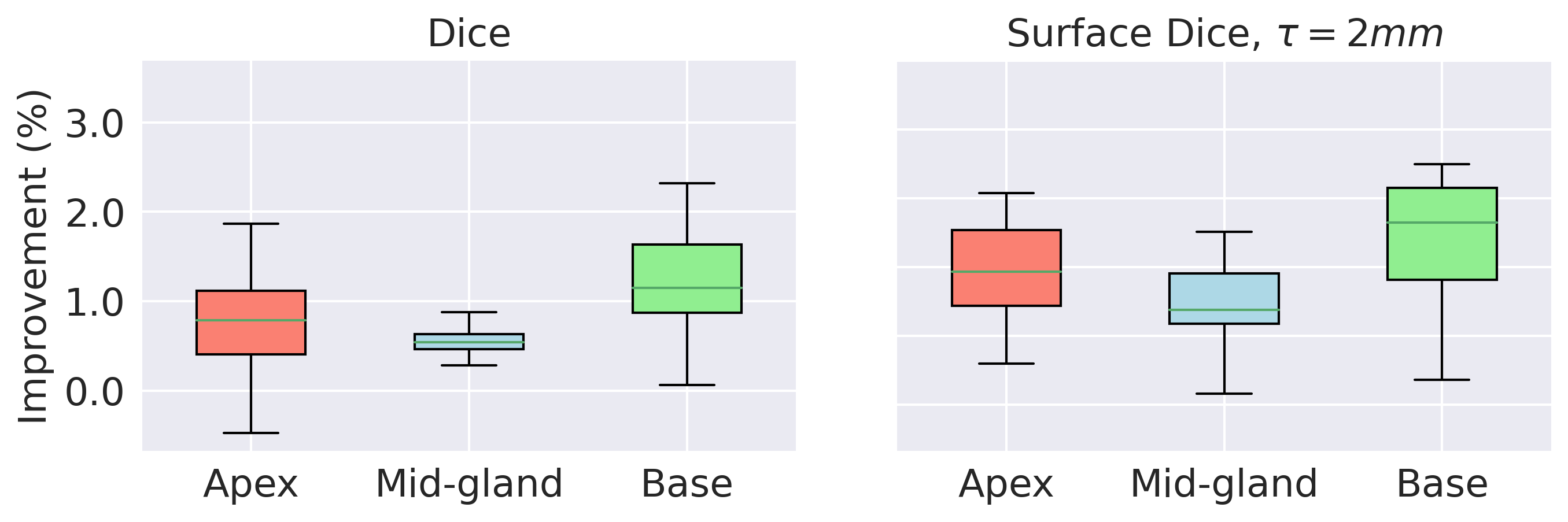}
    \caption{Improvement of the proposed approach for different prostate parts. Values are aggregated over all experimental setups with $N=50$ (two types of nets, three values of $\alpha$, three net initializations, see Section~\ref{sub:setups}). Dice and Surface Dice with $\tau=2mm$ performance metrics are shown.}
    \label{fig:variance}
\end{figure}

As shown in Figure~\ref{fig:variance}, the main improvement comes from the base of the prostate (on average, approximately $1.2\%$ in DSC, $2.6\%$ in SDSC with $\tau=2mm$). A smaller improvement is observed in the apex, while in the mid-gland part it is the most subtle. Naturally, our approach can result in larger improvements if the data variation is larger. 
As illustrated in Figure~\ref{fig:pictures}, the difference between the produced segmentation variants is largest in the base and apex parts. The quality of produced segmentations is the highest in the mid-gland part, however our approach leads to an improvement in segmentation quality in all parts of the prostate.
These results are in line with studies, e.g., \cite{montagne2021challenge} that show that the inter-observer variation is the largest in the base, smaller in the apex, and the smallest in the mid-gland.

We note that the improvement of our approach compared to a training procedure with no data partitioning is larger for larger $\alpha$. This means that producing more segmentation variants helps to make at least one of the variants closer to the reference segmentation though the data amount for training each of the decoders is smaller. 

\setlength\tabcolsep{2.5pt}
\begin{table}


\begin{tabularx}{\textwidth}{|c|X?X?X|X|X|X|X|X|X|X|X|}
\hline
& & & \multicolumn{3}{c|}{\textbf{No partitioning}} &  \multicolumn{3}{c|}{\textbf{\textbf{Our approach}}} & \multicolumn{3}{c|}{\textbf{Improvement (\%)}} \\ \hline
      Network & \textbf{$N$} & \textbf{$\alpha$}   &  \textbf{DSC}         &  \textbf{SDSC 2mm} &  \textbf{SDSC 4mm}         &  \textbf{DSC}         & \textbf{SDSC 2mm} &  \textbf{SDSC 4mm}           & \textbf{DSC}       &    \textbf{SDSC 2mm} &  \textbf{SDSC 4mm}          \\ \hline

\multirow{8}{*}{ResNet18-Unet} &  
50 &  1 & 0.863  & 0.670  & 0.917  &  -  &  -  &  -  &  -  &  -  &  - \\
& 50 &  2 & 0.864 & 0.675 & 0.920 & 0.871 & 0.701 & 0.931 & \textbf{0.88} & \textbf{3.92} & \textbf{1.15} \\
& 50 & 3 & 0.863 & 0.672 & 0.918 & 0.872 & 0.703 & 0.934 & \textbf{1.04} & \textbf{4.52} & \textbf{1.68} \\
& 50 & 4 & 0.865 & 0.677 & 0.922 & 0.874 & 0.711 & 0.936 & \textbf{1.14} & \textbf{5.14} & \textbf{1.58} 
 \\ \cline{2-12}
& 100 &  1 & 0.864  & 0.676  & 0.919  &  -  &  -  &  -  &  -  &  -  &  - \\
& 100 &  2 & 0.866  & 0.683  & 0.922  & 0.870  & 0.695  & 0.928  & \textbf{0.55}  & \textbf{1.87}  & \textbf{0.71} \\
& 100 &  3 & 0.867  & 0.686  & 0.923  & 0.872  & 0.702  & 0.930  & \textbf{0.67}  & \textbf{2.41}  & \textbf{0.83} \\
& 100 &  4 & 0.868  & 0.691  & 0.924  & 0.873  & 0.708  & 0.930  & \textbf{0.64}  & \textbf{2.51}  & \textbf{0.64} \\

\thickhline
    
\multirow{8}{*}{VGG19-Unet} 
 
& 50 &  1 & 0.868  & 0.683  & 0.924  &  -  &  -  &  -  &  -  &  -  &  - \\ 
& 50 &  2 & 0.868 & 0.686 & 0.926 & 0.876 & 0.713 & 0.938 & \textbf{0.95} & \textbf{3.94} & \textbf{1.24} \\
& 50 &  3 & 0.868 & 0.688 & 0.927 & 0.877 & 0.719 & 0.939 & \textbf{1.00} & \textbf{4.58} & \textbf{1.27} \\
& 50 &  4 & 0.870 & 0.692 & 0.929 & 0.878 & 0.723 & 0.942 & \textbf{1.02} & \textbf{4.58} & \textbf{1.38} 
\\ \cline{2-12}

& 100 &  1 & 0.871  & 0.700  & 0.929  &  -  &  -  &  -  &  -  &  -  &  - \\
& 100 &  2 & 0.872  & 0.701  & 0.933  & 0.874  & 0.708  & 0.936  & \textbf{0.31}  & \textbf{1.01}  & \textbf{0.29} \\
& 100 &  3 & 0.873  & 0.702  & 0.934  & 0.875  & 0.712  & 0.938  & \textbf{0.33}  & \textbf{1.38}  & \textbf{0.41} \\
& 100 &  4 & 0.871  & 0.699  & 0.932  & 0.875  & 0.718  & 0.937  & \textbf{0.47}  & \textbf{2.73}  & \textbf{0.62} \\

          \hline
          
\end{tabularx}
\caption {Main results. The "No partitioning" results denote a training procedure in which all network paths are trained with all samples from the training dataset (see Section~\ref{sub:setups}) with the number of paths equal to the corresponding $\alpha$ ($\alpha=1$ means that a single-path network is used). $N$ denotes the total number of patients in the training dataset (before partitioning). SDSC results are demonstrated for two tolerance values. DSC and SDSC scores are rounded to three decimal places.}

\label{tab:results}
\end{table}

\section{DISCUSSION}

 We applied the proposed segmentation approach to prostate segmentation on CT scans. However, our approach can be applied to other organs and image modalities as well. Moreover, it is not limited to binary segmentation and can be naturally extended to a multi-organ segmentation task by corresponding modification of the segmentation network and segmentation quality metric (to a multi-class one). 
 
In the results interpretation, we suggested that our algorithm managed to capture and utilize the observer segmentation variation present in the data. Such suggestion was supported by the fact that our algorithm showed the most substantial improvement in the base of the prostate - the same part which is known to have the largest inter-observer segmentation variation. However, it might be possible that there are other sources of variation in the data such as the presence of patients of different age groups, the use of varying imaging device settings, or different inclusion criteria of the seminal vesicles in the prostate delineated volume due to different degrees of prostate cancer spread into the seminal vesicles. Our algorithm does not distinguish between types of variation, and it only performs the partitioning to maximize segmentation quality. Making our approach insightful to end users (e.g., explicitly showing different types of variation in data) and validating the practical added value (i.e., reducing required time for scan delineation) is an interesting and important question for further research.

As mentioned in Section~\ref{sub:results}, a larger number of produced segmentation variants leads to better performance of our approach. However, for practical usage, a large number of segmentation variants might not be convenient as a medical expert needs time to select one of them. We suggest that the selection of $\alpha$ is done according to the user preferences as a trade-off between the diversity of segmentation variants and available time to assess them. Furthermore, we believe that a specialized graphical user interface needs to be developed and used in order to allow a medical expert to navigate through the suggested segmentation variants quickly and intuitively.

\section{CONCLUSIONS}
We have proposed a novel approach to medical image segmentation aimed at capturing variation in the data. Specifically, we proposed to achieve this by training a multi-path Unet-like segmentation network such that each path is trained on a specific data subset. The best partitioning into subsets is found automatically by solving an optimization problem using the surrogate-assisted evolutionary algorithm SA-P3-GOMEA.
In experiments with real clinical data consisting of CT scans with prostate segmentations, our approach was found to be able to detect the subsets of data automatically and produce multiple segmentation predictions resulting in a performance increase of auto-segmentation compared to not using data partitioning. Analysis of the achieved improvement showed that most improvement was achieved for the base of the prostate. This corresponds to the literature reported findings that the largest inter-observer segmentation variation occurs at the base of the prostate. This is indicative that our approach is indeed capable of capturing observer variation in prostate segmentation.

\acknowledgments 
This work is part of the research programme Commit2Data with project number
628.011.012, which is financed by the Dutch Research Council (NWO).

\bibliography{report} 

\begin{thebibliography}{10}

\bibitem{kohl2018probabilistic}
Kohl, S., Romera-Paredes, B., Meyer, C., De~Fauw, J., Ledsam, J.~R.,
  Maier-Hein, K., Eslami, S., Jimenez~Rezende, D., and Ronneberger, O., ``A
  probabilistic \relax{U-net} for segmentation of ambiguous images,'' {\em
  Advances in \relax{Neural Information Processing Systems}}~{\bf 31} (2018).

\bibitem{rupprecht2017learning}
Rupprecht, C., Laina, I., DiPietro, R., Baust, M., Tombari, F., Navab, N., and
  Hager, G.~D., ``Learning in an uncertain world: Representing ambiguity
  through multiple hypotheses,'' in [{\em Proceedings of the IEEE International
  Conference on Computer Vision}{\nolinebreak\hspace{0.1em}]},   3591--3600
  (2017).

\bibitem{dushatskiy2020observer}
Dushatskiy, A., Mendrik, A.~M., Bosman, P. A.~N., and Alderliesten, T.,
  ``Observer variation-aware medical image segmentation by combining deep
  learning and surrogate-assisted genetic algorithms,'' in [{\em Medical
  Imaging 2020: Image Processing}{\nolinebreak\hspace{0.1em}]},   {\bf 11313},
  113131B, International Society for Optics and Photonics (2020).

\bibitem{unet}
Ronneberger, O., Fischer, P., and Brox, T., ``\relax{U-Net}: Convolutional
  networks for biomedical image segmentation,'' in [{\em International
  Conference on Medical Image Computing and Computer-Assisted
  Intervention}{\nolinebreak\hspace{0.1em}]},   234--241, Springer (2015).

\bibitem{dushatskiy2021novel}
Dushatskiy, A., Alderliesten, T., and Bosman, P. A.~N., ``A novel
  surrogate-assisted evolutionary algorithm applied to partition-based ensemble
  learning,'' in [{\em Proceedings of the Genetic and Evolutionary Computation
  Conference}{\nolinebreak\hspace{0.1em}]},   583--591 (2021).

\bibitem{surfacedice}
Nikolov, S., Blackwell, S., Zverovitch, A., Mendes, R., Livne, M., De~Fauw, J.,
  Patel, Y., Meyer, C., Askham, H., Romera-Paredes, B., et~al., ``Deep learning
  to achieve clinically applicable segmentation of head and neck anatomy for
  radiotherapy,'' {\em arXiv preprint arXiv:1809.04430v3}  (2018).

\bibitem{Yakubovskiy:2019}
Yakubovskiy, P., ``{Segmentation} {Models} {Pytorch},'' {\em GitHub
  repository}~{\bf \url{https://github.com/qubvel/segmentation_models.pytorch}}
  (2020).

\bibitem{he2016deep}
He, K., Zhang, X., Ren, S., and Sun, J., ``Deep residual learning for image
  recognition,'' in [{\em \relax{Proceedings of the IEEE Conference on Computer
  Vision and Pattern Recognition}}{\nolinebreak\hspace{0.1em}]},   770--778
  (2016).

\bibitem{SimonyanZ14a}
Simonyan, K. and Zisserman, A., ``Very deep convolutional networks for
  large-scale image recognition,'' in [{\em 3rd International Conference on
  Learning Representations, {ICLR} 2015, San Diego, CA, USA, May 7-9, 2015,
  Conference Track Proceedings}{\nolinebreak\hspace{0.1em}]},  Bengio, Y. and
  LeCun, Y., eds. (2015).

\bibitem{montagne2021challenge}
Montagne, S., Hamzaoui, D., Allera, A., Ezziane, M., Luzurier, A., Quint, R.,
  Kalai, M., Ayache, N., Delingette, H., and Renard-Penna, R., ``Challenge of
  prostate {MRI} segmentation on {T2}-weighted images: inter-observer
  variability and impact of prostate morphology,'' {\em Insights into
  \relax{Imaging}}~{\bf 12}(1),  71 (2021).

\end{thebibliography}
\bibliographystyle{spiebib} 

\end{document}